\documentclass[prd,preprint,tightenlines,floatfix,showpacs,preprintnumbers,nofootinbib,eqsecnum]{revtex4}

 \usepackage[dvips,final]{graphicx}
  \usepackage{amssymb}
   \usepackage{amsmath} % Advanced math typesetting
    \usepackage{amsfonts}
     \usepackage{epsfig}
     \usepackage{color}
      \usepackage{bm} % bold math

\usepackage{booktabs}
\usepackage{multirow}
\usepackage{slashed}
\usepackage{comment}

 \newcommand\ra{\rangle}
 \newcommand\beq{\begin{equation}}
 
 \newcommand\eeq{\end{equation}}
 \newcommand\beqn{\begin{eqnarray}}
 \newcommand\eeqn{\end{eqnarray}}

 \def\gsim{\mathrel{\rlap{\lower4pt\hbox{\hskip1pt$\sim$}}
 \raise1pt\hbox{$>$}}}

 \def\fm{\,\mbox{fm}}
 
 \def\GeV{\,\mbox{GeV}}
 
 \def\lsim{\mathrel{\rlap{\lower4pt\hbox{\hskip1pt$\sim$}}
     \raise1pt\hbox{$<$}}}         %less than or approx. symbol
 \def\gsim{\mathrel{\rlap{\lower4pt\hbox{\hskip1pt$\sim$}}
     \raise1pt\hbox{$>$}}}         %greater than or approx. symbol

 \def\beq{\begin{equation}}
 \def\eeq{\end{equation}}
 
 \def\beqy{\begin{eqnarray}} 
 \def\eeqy{\end{eqnarray}}
 \def\dfr{\mathrm{d}}
\def\Jpsi{J\!/\!\psi}
\def\psip{\psi'}
\def\Y{\Upsilon}
\def\Yp{\Upsilon'}
\def\Ypp{\Upsilon''}
\begin{document}
%%%%%%%%%%%%%%%%%%%%%%%%

%======================================
\title
{
\vspace*{-1.00cm}
$D$-wave effects in diffractive electroproduction of heavy quarkonia 
from the photon-like $V\rightarrow Q\bar Q$ transition
}
%======================================

\author{Michal Krelina$^{1,2}$}
\email{michal.krelina@usm.cl}

\author{Jan Nemchik$^{2,3}$}
\email{nemcik@saske.sk}

\author{Roman Pasechnik$^{4,5,6}$}
\email{roman.pasechnik@thep.lu.se}

\affiliation{
{$^1$\sl
Departamento de F\'{\i}sica,
Universidad T\'ecnica Federico Santa Mar\'{\i}a,
Casilla 110-V, Valpara\'{\i}so, Chile
}\vspace{0.1cm}\\
{$^2$\sl 
Czech Technical University in Prague, FNSPE, B\v rehov\'a 7, 11519 
Prague, Czech Republic
}\vspace{0.1cm}\\
{$^3$\sl
Institute of Experimental Physics SAS, Watsonova 47, 04001 Ko\v 
sice, Slovakia
}\vspace{0.1cm}\\
{$^4$\sl
Department of Astronomy and Theoretical Physics, Lund
University, SE-223 62 Lund, Sweden
}\vspace{0.1cm}\\
{$^5$\sl
Nuclear Physics Institute ASCR, 25068 \v{R}e\v{z}, Czech Republic
}\vspace{0.1cm}\\
{$^6$\sl Departamento de F\'isica, CFM, Universidade Federal 
de Santa Catarina, C.P. 476, CEP 88.040-900, Florian\'opolis, 
SC, Brazil
%\vspace{3mm}
}
}

%=========================================
\begin{abstract}
\vspace{2mm}
We analyse the validity of a commonly used identification between structures of the virtual photon $\gamma^*\rightarrow Q\bar Q$ and vector meson $V\rightarrow Q\bar Q$ transitions. In the existing studies of $S$-wave vector-meson photoproduction in the literature, such an identification is typically performed in the light-front (LF) frame while the radial component of the meson wave function is rather postulated than computed from the first principles. The massive photon-like $V\rightarrow Q\bar Q$ vertex, besides the $S$-wave component, also contains an extra $D$-wave admixture in the $Q\bar Q$ rest frame. However, the relative weight of these contributions cannot be justified by any reasonable nonrelativistic $Q\bar Q$ potential model. In this work, we investigate the relative role of the $D$-wave contribution starting from the photon-like quarkonium $V\rightarrow Q\bar Q$ transition in both frames: in the $Q\bar Q$ rest frame (with subsequent Melosh spin transform to the LF frame) and in the LF frame (without Melosh transform). In both cases, we employ the same radial wave functions found in the potential approach. 
We show that the photon-like transition imposed in the $Q\bar Q$ rest frame, with subsequent boosting to the LF frame, leads to significant discrepancies with the experimental data. In the second case, i.e. imposing the photon-like transition straight in the LF frame, we find that the corresponding total $\Jpsi(1S)$ photoproduction cross sections are very close to those obtained with the ``$S$-wave only'' $V\rightarrow Q\bar Q$ transition, both leading to a good description of the data. However, we find that the ``$S$-wave only'' transition leads to a better description of photoproduction data for excited (e.g. $\psip(2S)$) heavy quarkonium states, which represent a more effective tool for study of $D$-wave effects. Consequently,
the predictions for production of excited states based on the photon-like structure of $V\rightarrow Q\bar Q$ transition should be treated with a great care due to a much stronger sensitivity of the $D$-wave contribution to the nodal structure of quarkonium wave functions. 
\end{abstract}
%=================================================================================


\maketitle

%
%
%
%======================
\section{Introduction}
\label{Sec:Intro}
%======================
%
%
%

Elastic virtual photo- and electroproduction of heavy quarkonia $\gamma^*\,p\to\Jpsi(\psip,\Y,\Yp,...)\,p$ remains one of the most active fundamental research areas of Quantum Chromo-Dynamics (QCD). These processes provide a very powerful tool for studies of various QCD phenomena in the framework of perturbative QCD (pQCD) since the length scale associated with heavy quarkonia is relatively small. This fact allows to minimize the underlined QCD uncertainties that emerge due to nonperturbative interactions (see Refs.~\cite{sr-18,sr-19}, while a comprehensive review on quarkonia phenomenology can be found in e.g.~Refs.~\cite{ivanov-04,brambilla-04,brambilla-10}). On the other hand, in nucleus-nucleus collisions the production of heavy charmonia and bottomonia represent an important probe for the hot medium that emerges in heavy-ion collisions (see Ref.~\cite{satz}, for example). 

The powerful method for probing the pQCD contribution to the quarkonium production amplitude is based on the so-called \textit{scanning phenomenon} \cite{bzk-93,bzk-94,jan-94,jan-96,jan-01,jan-07}. This phenomenon has been previously studied in detail in Ref.~\cite{sr-18} in the case of the photo- and electroproduction of heavy quarkonia. Here, the corresponding production amplitude is scanned at dipole sizes $r \sim r_S$, where the scanning radius $r_S$ is defined as follows
%
%========================================
\beqn
r_S \approx  
\frac{Y}{2\sqrt{Q^{2}\,z\,(1-z) + m_Q^{2}}}
\sim
\frac{Y}{\sqrt{Q^2 + M_V^2}}\, .
\label{scan-rad}
\eeqn
%========================================
%
Here, $M_V$ is the mass of a heavy quarkonium state $V$, $m_Q$ is the heavy quark mass, $Q^2$ is the photon
virtuality, and $z$ is the longitudinal momentum fraction of the photon taken by a quark (antiquark) in the photon leading-order Fock $\gamma^\star \to Q\bar Q$ fluctuation. The latter represents a dipole of transverse separation $\vec r$. The factor $Y\sim 6$ in the nonrelativistic limit ($z\sim 1/2$, $M_V\sim 2 m_Q$) as was previously found in Refs.~\cite{jan-94,sr-18}.

According to the scanning phenomenon, the onset of pQCD is reached at sufficiently small values of $r_S\lsim r_0$, where $r_0\sim 0.3\,\fm$ is the so-called gluon propagation radius \cite{spots,drops}. In the case of heavy quarkonia production, this condition provides \cite{jan-07,sr-18},
%
%----------------------------------------
 \beqn
(Q^2 + M_V^2)\gsim Q_{\rm pQCD}^2 = \frac{Y^2}{r_0^2}\, ,
\label{pqcd}
\eeqn
%----------------------------------------
%
which is satisfied only for the photo- and electroproduction of bottomonia. However, in the case of charmonium production one requires relatively large photon virtualities, typically $Q^2\gsim 10\div 20\,\GeV^2$, in order to probe the pQCD regime. 

Another manifestation of the scanning phenomenon is based on the universality properties in production of different heavy quarkonium states as a function of the scaling variable $Q^2 + M_V^2$ \cite{jan-94,jan-96,jan-07,sr-18}. 

So, elastic electroproduction of heavy quarkonia at $Q^2\gg 0$ is an efficient tool for studying the QCD phenomena by probing physics mainly the hard scale which enables to minimize theoretical uncertainties and to safely rely on pQCD approach in the analysis of the corresponding observables \cite{sr-19}. On the other hand, only charmonium electroproduction at small $Q^2\lsim M_V^2$ (see Eq.~(\ref{pqcd})) probes the semihard-scale physics mainly of nonperturbative origin. 

Consequently, in Refs.~\cite{sr-18,sr-19} namely the marginal theoretical uncertainties in description of heavy quarkonium electroproduction represented the keystone for a detailed analysis of the \textit{Melosh spin rotation} effects \cite{melosh} performed in the framework of color dipole formalism  \cite{bzk-91,bzk-93,bzk-94,jan-94,jan-94-2s,jan-96,yura-00,jan-00a,jan-00b}. As a result, a strong onset of these effects was found significantly modifying the quarkonium production cross sections, especially for radially-excited $2S$ and $3S$ production. 

The corresponding analysis of spin effects has been performed \cite{sr-18,sr-19} assuming the Light-Front (LF) wave functions of $Q\bar Q$ fluctuation for heavy quarkonium states with only $S$-wave structure\footnote{The LF wave-function formalism accounting for the Melosh spin transformation has also been recently applied in studies of $\gamma^*\gamma^*\to \eta_c(1S),\eta_c(2S)$ transition form factor in Ref.~\cite{Babiarz:2019sfa}}. However, the existing studies of charmonium production in the literature employ an unjustified assumption about a photon-like structure of the LF quarkonium wave function which contains the contributions of both $S$- and $D$-wave states in the $c\bar c$ rest frame. Consequently, the weight of the $D$-wave state in the latter approach is strongly correlated with the structure of the charmonium vertex and cannot be understood in the framework of any realistic nonrelativistic $c-\bar c$ interaction potential~\cite{Ivanov:1999wv}. 

In principle, the massive photon-like vector-meson $V\rightarrow Q\bar Q$ transition can be imposed either in the $Q\bar Q$ rest frame, such that a subsequent boosting of the spin-dependent and radial components to the LF (or infinite-momentum) frame has to be performed, or right in the LF frame as is typically done in the literature. In the former case, the Melosh spin transform is necessary while it is assumed to be effectively included in the latter case. In the present paper, we analyze separately both cases since any of them is not fully justified in the literature.

In the first case, we compute a modification of quarkonium production yields caused by the spin rotation effects and derive for the first time the corresponding formulas for electroproduction cross sections, while the radial counterpart is found in the potential approach starting from a selection of realistic interaction potentials. The numerical predictions obtained with such virtual photon-like structure of the quarkonium vertex are compared with our previous results \cite{sr-18,sr-19} where the $V\to Q\bar Q$ transition was derived started from the ``$S$-wave only'' structure in the $Q\bar Q$ rest frame. As one of our key results, we find a strong correlation between the onset of spin effects and the structure of the distribution amplitude of $Q\bar Q$ fluctuation for the heavy quarkonium. The presence of a $D$-wave admixture, combined with the Melosh transform, leads to a strong disagreement with the charmonium photoproduction data, even for the ground $\Jpsi(1S)$ state. 

In the second case when no Melosh spin rotation is performed, while the radial component is still treated in the framework of the Schroedinger equation, the $\Jpsi(1S)$ photoproduction cross section appears to be very close to that obtained in the framework of ``$S$-wave only'' structure in the $Q\bar Q$ rest frame together with spin effects. This means that the $D$-wave component in the LF photon-like $V\to Q\bar Q$ transition has quantitatively a very similar impact on the observables as the Melosh transform acted upon ``$S$-wave only'' spin-dependent structure. 

However, we find that such similarities in predictions based on the ``$S$-wave only'' structure in the $Q\bar Q$ rest frame and the popular photon-like transition in the LF frame are destroyed for production of the excited quarkonium states. In particular, we demonstrate that the onset of the $D$-wave effect, followed from the photon-like heavy quarkonium vertex, appears to be stronger in production of the radially-excited $2S$ and $3S$ states compared to the $1S$ state production as a direct manifestation of the nodal structure in the corresponding radial wave functions.

The paper is organised as follows. 
In Sec.~\ref{Sec:col-dip}, we present a short review of the factorized color dipole formalism and its application to elastic heavy quarkonium photo- and electroproduction. 
Sec.~\ref{Sec:results} is devoted to comparison of our model parameter-free calculations with available data on $\Jpsi(1S)$ and $\Y(1S)$ photoproduction. 
Here we also demonstrate the higher sensitivity of model predictions to the $D$-wave and Melosh spin rotation effects investigating the production of radially excited $2S$ and $3S$ heavy quarkonia. 
Finally, in Sec.~\ref{Sec:conclusions} we summarise the results obtained in our study.

%
%
%
%=====================================================
\section{An overview of the color-dipole approach 
to electroproduction of heavy quarkonia}
\label{Sec:col-dip}
%=====================================================
%
%
%

In the present paper, we treat the photo- and electroproduction of $S$-wave heavy quarkonium states in the framework of color dipole approach \cite{bzk-91,bzk-93,bzk-94,jan-94,jan-96,yura-00,sr-18,sr-19}.
In the target rest frame the projectile photon (real, with $Q^2\to 0$, or virtual) undergoes strong interactions via its Fock components with the target. Here, the lowest state in the Fock state expansion for the relativistic vector meson is represented by the quark-antiquark $Q\bar Q$ pair.
Such a state can be considered as a color dipole whose scattering depends on the transverse separation between the quark and antiquark and on the light-cone (LC) momentum fraction of the meson taken by a heavy (anti)quark. Consequently, the dipole interactions with a target are described by the dipole cross section $\sigma_{Q\bar Q}$ (introduced for the first time in Ref.~\cite{zkl}), which is a flavor-independent universal function of the dipole separation and the Bjorken $x$ (or the dipole-target scattering energy). Here, in the leading-$\log(1/x)$ approximation the energy dependence of the dipole cross section emerges due to the presence of the higher Fock states with radiated gluons, for example, $Q\bar Q+g$ etc.

We start with the expression for the elastic electroproduction amplitude in the forward limit in the rest frame of the target~\cite{bzk-91,bzk-93,bzk-94,jan-94,jan-96,yura-00},
%
%------------------------------------------
\beqn
\mathrm{Im}\,\mathcal{A}^{\gamma^{\ast}p\rightarrow Vp}_{T,L}(x,Q^2)
=
\int\dfr^{2}r\,\int\limits_{0}^{1}\dfr z \,\Phi^{\dagger}_{V_{T,L}}(r,z)\,
\Psi_{\gamma^{\ast}_{T,L}}(r,z;Q^2)\,\sigma_{Q\bar Q}(x,r)\,.
%...............
\label{prod-amp}
%...............
\eeqn
%------------------------------------------
%
Here, the function $\Psi_{\gamma^{\ast}_{T,L}}(r,z;Q^2)$ is the LF wave function of the photon with virtuality $Q^2$ and with a transverse (T) or longitudinal (L) polarisation corresponding to its Fock $Q\bar Q$ component; $\vec r$ represents the two-dimensional transverse separation within the $Q\bar Q$-dipole; the variable $z=p_Q^+/p_{\gamma}^+$ is the boost-invariant fraction of the photon momentum carried by a heavy quark (or antiquark); $\Phi_{V_{T,L}}(r,z)$ is the LF vector meson wave function for $T$ and $L$ polarized heavy quarkonia sharing exactly the same LC variables with the photon wave functions; $\sigma_{Q\bar Q}(x,r)$ describes an interaction of the $Q\bar Q$ dipole (with transverse separation $r\equiv |\vec{r}|$) with the target, where $x=(M_V^2+Q^2)/s$ which coincides with the standard Bjorken variable at large $Q^2\gg M_V^2$, with the c.m. energy squared $s$ of the electron-proton system (for more details, see Ref.~\cite{ryskin-95}). One safely employs the nonrelativistic QCD (NRQCD) limit in studies of heavy quarkonia production in the relevant (measured) kinematical domains. In the computation of the total electroproduction $\gamma^{\ast}p\rightarrow V p$ cross section, we follow Ref.~\cite{yura-00}, such that
%
%------------------------------------------------------
\beqn
\sigma^{\gamma^{\ast}p\rightarrow Vp}(x,Q^2)
=
\sigma^{\gamma^{\ast}p\rightarrow Vp}_{T} +
\tilde{\varepsilon}\,
\sigma^{\gamma^{\ast}p\rightarrow Vp}_{L}
=
\frac{1}{16\pi B}\left(\Big\vert 
\mathcal{A}^{\gamma^{\ast}p\rightarrow Vp}_{T}\Big\vert^{2}+
\tilde{\varepsilon}\Big\vert \mathcal{A}^{\gamma^{\ast}p\rightarrow Vp}_{L}
\Big\vert^{2}\right) \,.
\label{final}
\eeqn
%------------------------------------------------------
%
Here, $B=4.73$ GeV$^{-2}$ is the slope parameter and $\tilde{\varepsilon} = 0.99$ represents the photon polarization (used also in Refs.~\cite{sr-18,sr-19}), both are independent of spin effects and are obtained by fitting to the H1 data at HERA \cite{h1-00}. In the analysis of the Melosh spin rotation effects and their interplay with the $D$-wave contribution to the quarkonium wave function, the energy and scale dependence of the elastic slope $B$ play a minor role and hence were neglected in the rest of this work. For a detailed analysis of the $W$ and $Q^2$ dependence of the elastic slope and their impact on the photoproduction observables, see Refs.~\cite{sr-18,sr-19}.

The small real-part corrections are usually incorporated into the corresponding amplitudes $\mathcal{A}_{T,L}$ as follows \cite{bronzan-74,jan-96,forshaw-03},
%
%------------------------------------------------------
\beqn
{\mathcal A}^{\gamma^{\ast}p\rightarrow Vp}_{T,L}(x,Q^2) 
\simeq
{\mathrm{Im}\mathcal{A}}^{\gamma^{\ast}p\rightarrow Vp}_{T,L}(x,Q^2)
\,
\left(1 - i\,\frac{\pi}{2}\,\frac
{\partial
 \,\ln\,{\mathrm{Im}\mathcal{A}}^{\gamma^{\ast}p\rightarrow Vp}_{T,L}(x,Q^2)}
{\partial\,\ln x} \right)\, .
%--------------
  \label{re/im}
%--------------
\eeqn
%------------------------------------------------------
%
We have cross-checked numerically that the difference between this approximate formula 
and the exact expression for the real-to-imaginary ratio is relatively small 
(at the level of a few percent) in the kinematical domains of our analysis. So,
for practical purposes of the current work it is sufficient to use the approximated 
expression (\ref{re/im}).

The LF photon wave function, corresponding to the perturbative (leading-order) Fock 
$\gamma^\star_{L,T} \to Q\bar Q$ fluctuations reads \cite{kogut,bjorken,nnn}, 
%
%------------------------------------------------------
\beqn
\Psi^{(\mu,\bar\mu)}_{\gamma^{\ast}_{T,L}}(r,z;Q^2)
=
\frac{\sqrt{N_c\alpha_{\rm em}}}{2\pi} Z_Q\,\chi_Q^{\mu\dag}
\hat{\mathcal{O}}^{T,L}_{\gamma^{\ast}}\tilde\chi_{\bar Q}^{\mu}\,K_0(\varepsilon r) \,.
\label{gam-wf}
\eeqn
%------------------------------------------------------
%
As usual, here we denote the number of QCD colors, $N_c = 3$, charge of the heavy quark, $Z_Q$ (for the relevant cases considered in this work, $Z_c=2/3$ and $Z_b = 1/3$, for the charm and bottom quarks, respectively), $\varepsilon^2=z(1-z)Q^2+m_Q^2$, the modified Bessel function of the second kind, $K_0(\varepsilon r)$, while the two-component (anti)quark spinors in the LF frame $\chi^\mu_Q$ and $\tilde\chi^{\bar\mu}_{\bar Q}\equiv i\sigma_y\chi^{\bar\mu\ast}_{\bar Q}$ normalized as (see e.g.~Ref.~\cite{bzk-dy-01})
%
%------------------------------------------------------
\beqn
\sum\limits_{\mu,\bar \mu}\left(\chi^{\mu\dag}_Q\hat A
\tilde\chi^{\bar\mu}_{\bar Q}\right)^\ast\left(\chi^{\mu\dag}_Q\hat B
\tilde\chi^{\bar\mu}_{\bar Q}\right)
=
\mathrm{Tr}(\hat A^\dag\hat B)\, .
\eeqn
%------------------------------------------------------
%

Besides, in Eq.~(\ref{gam-wf}) we introduce the operators in the spinor space denoted as $\hat{\mathcal{O}}^{T,L}_{\gamma^{\ast}}$ which are represented in the form
%
%------------------------------------------------------
\beqn
\hat{\mathcal{O}}^T_{\gamma^{\ast}}
&=& 
m_Q\,\vec\sigma\cdot\vec e_\gamma + 
i(1-2z)(\vec\sigma\cdot\vec n)(\vec e_\gamma\cdot\vec\nabla_r) +
(\vec n\times\vec e_\gamma)\cdot\vec\nabla_r \,, \nonumber\\
\hat{\mathcal{O}}^L_{\gamma^{\ast}}
&=& 
2\,Q\,z(1-z)\vec\sigma\cdot \vec n \,, \qquad \vec\sigma
=
(\sigma_x,\sigma_y,\sigma_z) \,, \qquad \vec\nabla_r 
\equiv \partial/\partial \vec{r} \,,
\label{o-gam}
\eeqn
%------------------------------------------------------
%
written in terms of the Pauli matrices, $\sigma_i$, $i={x,y,z}$, the photon transverse polarisation, $\vec e_\gamma$, and the unit vector directed along the momentum of the initial photon $\vec n=\vec p_{\gamma}/|\vec p_{\gamma}\,|$.

In our previous studies \cite{sr-18,sr-19} (see also Ref.~\cite{yura-00}) we assumed a simple $S$-wave structure of the vertex $V \to Q\bar Q$ corresponding to only the first non-derivative part of expression for $\hat{\mathcal{O}}^T_{\gamma^{\ast}}$ in Eq.~(\ref{o-gam}). Among the other two terms in $\hat{\mathcal{O}}^T_{\gamma^{\ast}}$ that contain spatial derivative, $\vec\nabla_r$, only the last one survives in a vicinity of equal momentum sharing between $Q$ and $\bar Q$, $z=1/2$. 

If one adopts the massive photon-like $V\rightarrow Q\bar Q$ transition in the $Q\bar Q$ rest frame, then in general, besides the $S$-wave, it should contain also a $D$-wave contribution. Indeed, while the first term in $\hat{\mathcal{O}}^T_{\gamma^{\ast}}$ is a well-known $S$-wave term \cite{yura-00}, the last (derivative) term would effectively represent the $D$-wave contribution in the meson rest frame. A priori there is no conclusive way to justify the relative weight of the $D$-wave admixture, namely, it cannot be straightforwardly obtained in the potential approach (i.e. by using any well-defined realistic $Q\bar Q$ interaction potential in the Schroedinger equation). Besides, it is strongly correlated with the structure of the quarkonium $V\rightarrow Q\bar Q$ vertex. Here, we follow a naive approach using the same radial wave function as in our earlier ``$S$-wave only'' analysis in Refs.~\cite{sr-18,sr-19} and apply the Melosh spin transform acting on the rest-frame photon-like $V \to Q\bar Q$ vertex operator in order to convert it to the LF frame. The results will then be compared to a more conventional approach relying on an unjustified assumption of the photon-like structure of the vector-meson wave function imposed directly in the LF frame.

Consequently, we start with the wave function for the $Q\bar Q$ Fock component of $S$-wave heavy $T$ and $L$ polarized quarkonia in Eq.~(\ref{prod-amp}) allowing the following factorization of the spatial and spin-dependent parts,
%
%------------------------------------------------------
\beqn
\Phi^{(\mu,\bar\mu)}_{V_{T,L}}(r,z)
=
\frac{\sqrt{N_c}}{\sqrt{2}} \,\chi_Q^{\mu\dag}
\hat{\mathcal{O}}^{T,L}_V\tilde\chi_{\bar Q}^{\mu}\,\Psi_{V_{T,L}}(r,z) \,,
 \qquad 
\Psi_{V_{T,L}}(r,z) = {\mathcal N}_{T,L} \Psi_{V}(r,z) \,,
\label{vm-wf}
\eeqn
%------------------------------------------------------
%
where ${\mathcal N}_{T,L}$ are the normalisation factors
satisfying
%
%----------------------------------------------------
\beqn
N_c \int\dfr^2 r \int\dfr z \bigl\{ m_Q^2\,|\Psi_{V_T} (\vec r,z)|^2 + [z^2+(1-z)^2]\,
|\partial_r  \Psi_{V_T}(\vec r,z) |^2 \bigr\} = 1\\
4\,N_c\,M_V^2 \int\dfr^2 r \int\dfr z\,\,z^2 (1-z)^2  |\Psi_{V_L} (\vec r,z)|^2 = 1 \,,
\eeqn
%-----------------------------------------------------
%
and the operators $\hat{\mathcal{O}}^{T,L}_V$ take a similar form as those in Eq.~(\ref{o-gam}),
%
%------------------------------------------------------
\beqn
\hat{\mathcal{O}}^T_V
&=& 
m_Q\,\vec\sigma\cdot\vec e_V + 
i(1-2z)(\vec\sigma\cdot\vec n)(\vec e_V\cdot\vec\nabla_r) +
(\vec n\times\vec e_V)\cdot\vec\nabla_r \,, \label{o-vm-T}\\
\hat{\mathcal{O}}^L_V
&=& 
2\,M_V\,z(1-z)\vec\sigma\cdot \vec n \,.
\label{o-vm-L}
\eeqn
%------------------------------------------------------
%
Here, $\vec e_{V}$ is the vector meson polarization vector. Such a form of the heavy photon-like $V\to Q\bar Q$ wave function has been initially postulated in the LF frame by a convention in the literature~\cite{ryskin-92,brodsky-94,frankfurt-95,jan-96}. For further references, let us denote this case as {\it scenario I}. Here the two-component spinors $\chi_Q^{\mu\dag}$ and $\tilde\chi_{\bar Q}^{\mu}$ in Eq.~(\ref{vm-wf}) are readily assumed to be defined in the LF frame, thus no additional Melosh spin transformation is required in this case. However, such a wave function appears to be a mixture of $S$- and $D$-wave components with unknown weights which cannot be easily disentangled the $Q\bar Q$ rest frame.

An alternative way denoted in what follows as {\it scenario II} would rely on imposing the massive photon-like structure of the quarkonium wave function (\ref{vm-wf}) in the $Q\bar Q$ rest frame instead. This possibility is suggested by a particularly simple form of $\hat{\mathcal{O}}^T_V$ where the ``$S$-wave only'' component is readily given only by the first (non-derivative) term in Eq.~(\ref{o-vm-T}) studied previously in our earlier works \cite{sr-18,sr-19} (and will be referred to here as {\it scenario III}), while the second and third terms in Eq.~(\ref{o-vm-T}) contain the spatial derivative, $\vec\nabla_r$, and would therefore correspond to a $D$-wave component. Here while the second term vanishes in the case of equal energy sharing between the heavy quark and antiquark, the third term dominates the $D$-term effect in this kinematics, and thus it is instructive to determine how strong is the corresponding impact on the quarkonium photoproduction observables. Assuming such a simple structure of the $D$-wave component and a clear $S$-$D$ wave separation, the radial wave function $\Psi_{V}(r,z)$ in the $Q\bar Q$ rest frame can then be consistently found in the potential approach based upon the Schroedinger equation following the procedure developed in Refs.~\cite{sr-18,sr-19}. 
Of course, the $D$-wave component is effectively present in both scenarios I and II, and its interplay with the spin rotation effects remains unclear.

In the latter scenario II, one expects to employ the Melosh spin transform and to effectively rotate the quark and antiquark spinors in Eq.~(\ref{vm-wf}) from the meson rest frame to the LF frame. The spin rotation causing a transformation $\Phi^{(\mu,\bar\mu)}_{V} \to \tilde{\Phi}^{(\mu,\bar\mu)}_{V}$ can be straightforwardly incorporated as follows. Starting from the factorised expression
%
%------------------------------------------------------
\beqn
\tilde{\Phi}_{V_{T,L}}^{(\mu,\bar\mu)}(\vec p_T,z)
=
\sqrt{N_c}\,
U^{(\mu,\bar\mu)}_{T,L}(\vec p_T,z)\Psi_{V_{T,L}}(p_T,z) \,,
%............
\label{PsiQQ}
%............
\eeqn
%------------------------------------------------------
%
where $\Psi_{V}(p_T,z)$ is the Fourier-image of the spatial part of the quarkonium wave 
function $\Psi_{V}(r,z)$ introduced in Eq.~(\ref{vm-wf}), such that
%
%----------------------------------------------------
\beqn
\Psi_{V}(r,z)
=
\int\limits_0^{\infty}\dfr p_T\, p_T J_0(p_T\,r)\,\Psi_V(p_T,z)\,,
\label{PsiV_r-to-pT}
\eeqn
%----------------------------------------------------
%
and the operators
%
%------------------------------------------------------
\beqn
U^{(\mu,\bar\mu)}_{T,L}(\vec p_T,z)
=
\frac{1}{\sqrt{2}}\xi_Q^{\mu\dag}\,\hat{\mathcal{O}}^{T,L}_V
\tilde\xi_{\bar Q}^{\bar \mu}\,,\qquad 
\tilde\xi_{\bar Q}^{\bar\mu}
=
i\sigma_y\xi_{\bar Q}^{\bar\mu\ast} \,, 
%...........
\label{Uini}
%...........
\eeqn
%------------------------------------------------------
%
are written in the vector-meson rest frame in terms of quark spinors $\xi$ connected 
to spinors $\chi$ in the LF frame by the following relation,
%
%------------------------------------------------------
\beqn
\xi^\mu_Q
=
R(\vec p_T,z)\chi_Q^\mu \,, 
\qquad \xi_{\bar Q}^{\bar\mu}
=
R(-\vec p_T,1-z)\chi_{\bar Q}^{\bar\mu} \,.
%...........
\label{spinrot}
%...........
\eeqn
%------------------------------------------------------
%
This transformation has a form of rotation in the space of spinor indices and is well-known as 
the Melosh spin transform (for more details, see Refs.~\cite{melosh,yura-00}). Here, the rotation 
matrix $R$ is found as
%
%------------------------------------------------------
\beqn
R(\vec p_T,z)
=
\frac{m_Q + z M - i(\vec\sigma\times\vec n)\cdot\vec p_T}
{\sqrt{(m_Q+z\,M)^2+p_T^2}} \,,
%...........
\label{melosh}
%...........
\eeqn
%------------------------------------------------------
%
where the effective mass of the $Q\bar Q$ pair,
expressed in terms of light-cone variables, is given by  
$M^2 = (p_T^2 + m_Q^2)/z(1-z)$. 
Here the limit of no spin transformation, corresponding to the unit $R$-matrix, 
leads to a reduction of the expression Eq.~(\ref{PsiQQ}) 
to Eq.~(\ref{vm-wf}) (i.e. $\tilde{\Phi} \to \Phi$).

Substituting
%
%------------------------------------------------------
\beqn
\tilde\xi_{\bar Q}^{\bar\mu}
=
i\sigma_yR^\ast(-\vec p_T,1-z)(-i)\sigma_y^{-1}
\tilde\chi_{\bar Q}^{\bar\mu} \,, \qquad
\xi_Q^{\mu\dag}=\chi_Q^{\mu\dag}R^{\dag}(\vec p_T,z) \,,
\eeqn
%------------------------------------------------------
%
into Eq.~(\ref{Uini}) one obtains
%
%------------------------------------------------------
\beqn
U^{(\mu,\bar\mu)}_{T,L}(\vec p_T,z)
=
\frac{1}{\sqrt{2}}\chi_Q^{\mu\dag}R^\dag(\vec p_T,z)\,
\hat{\mathcal{O}}^{T,L}_V\,\sigma_y R^\ast (-\vec p_T,1-z)
\sigma_y^{-1}\tilde\chi_{\bar Q}^{\bar \mu} \,.
\eeqn
%------------------------------------------------------
%
Consequently, the resulting dipole formula for the quarkonium electroproduction amplitude reads,
%
%------------------------------------------------------
\beqn
\mathrm{Im}\mathcal{A}^{\gamma^{\ast}p\rightarrow Vp}_{T,L}(x,Q^2)
=
\int\limits_0^1\dfr z \int\dfr^2r \Sigma_{T,L}(\vec r,z;Q^2)\,
\sigma_{Q\bar Q}(x,r)\, ,
\label{sr1}
\eeqn
%------------------------------------------------------
%
where
%
%------------------------------------------------------
\beqn
\Sigma_{T,L}(\vec r,z;Q^2)
=
\int\frac{\dfr^2 p_T}{2\pi}e^{-i\vec p_T\vec r}
\Psi_{V_{T,L}}(p_T,z)\sum\limits_{\mu,\bar\mu}U^{\dagger(\mu,\bar\mu)}_{T,L}(\vec p_T,z)
\Psi^{(\mu,\bar\mu)}_{\gamma^{\ast}_{T,L}}(r,z;Q^2)\,.
\label{sr2}
\eeqn
%------------------------------------------------------
%

Finally, and taking into account the no spin-flip contribution only, together with Eqs.~(\ref{sr1}) and (\ref{sr2}),
one arrives at the following final expressions,
%
%------------------------------------------------------
\beqn
&& \mathrm{Im}\mathcal{A}_{L}(x,Q^2)
=
Z_Q\,\frac{N_c\,\sqrt{\alpha_{em}}}{2\pi\sqrt{2}}\,
\int\limits_0^1\dfr z\int\dfr^2\,r\,\Sigma_{L}(\vec r,z;Q^2)\,
\sigma_{Q\bar Q}(x,r) \,,  
 \nonumber
\\
&& \Sigma_L
= 
8\,M_V\,Q\,z^2 (1-z)^2 K_0 (\varepsilon r)\int \dfr p_T\,p_T\,J_0(p_T\,r)
\Psi_{V_L}(p_T,z)\frac{m_T m_L + m_Q^2}{m_Q(m_T+m_L)} 
\,,
%.........
\label{AL}
%......... 
\eeqn
%-----------------------------------------------------
%
for a longitudinally polarized photon and quarkonium, and
%
%-----------------------------------------------------
\beqn
&& \mathrm{Im}\mathcal{A}_{T}(x,Q^2)
=
Z_Q\,\frac{N_c\,\sqrt{\alpha_{em}}}{2\pi\sqrt{2}}\,
\int\limits_0^1\dfr z\int\dfr^2r \,
\sigma_{Q\bar Q}(x,r)
\left[\Sigma^{(1)}_{T}(\vec r,z;Q^2) 
+
\Sigma^{(2)}_{T}(\vec r,z;Q^2)\right]\,,   
 \nonumber
\\
&&\Sigma^{(1)}_{T} 
= 
K_0(\varepsilon r)\,
\int \dfr p_T\,p_T\,J_0(p_T r)\Psi_{V_T}(p_T,z)\,
\frac{m_Q^2 (m_L^2 + 2 m_L m_T + 2 m_T^2) - m_L^2 m_T^2}{m_Q(m_T+m_L)} 
\,, \nonumber \\
&& \Sigma^{(2)}_{T} 
= 
K_1(\varepsilon r)\,\varepsilon\,
\int \dfr p_T\,p_T^2\,J_1(p_T r)\Psi_{V_T}(p_T,z)\,
\frac{m_L m_T (m_Q^2 - m_L^2)}{m_Q^3(m_T+m_L)} 
\,,
%.........
\label{AT}
%.........  
\eeqn
%------------------------------------------------------
%
for a transversely polarised photon and quarkonium, where
%
%------------------------------------------------------
\beqn
m_T^2 = m_Q^2 + p_T^2 \,, 
\qquad m_L^2 = 4m_Q^2\,z(1-z) \,.
\eeqn
%------------------------------------------------------

In the conventional scenario I, i.e. without imposing an extra Melosh transform in Eq.~(\ref{vm-wf}), the corresponding expressions for the $\Sigma_{L}$ and $\Sigma_T^{(1,2)}$ functions in Eqs.~(\ref{AL}) and (\ref{AT}) are reduced to the following simple and standard form,
%
%------------------------------------------------------
\beqn
\Sigma_{L}(\vec r,z;Q^2)
= 
8\,M_V\,Q\,z^2 (1-z)^2 K_0 (\varepsilon r)\int \dfr p_T\,p_T\,J_0(p_T\,r)
\Psi_{V_L}(p_T,z)\,,
\eeqn
%-----------------------------------------------------
%
and 
%
%-----------------------------------------------------
\beqn
\Sigma^{(1)}_{T}(\vec r,z;Q^2) 
= 2\,m_Q^2\,
K_0(\varepsilon r)\,
\int \dfr p_T\,p_T\,J_0(p_T r)\Psi_{V_T}(p_T,z)\,, 
\nonumber \\
\Sigma^{(2)}_{T}(\vec r,z;Q^2) 
= 
2 \bigl [z^2 + (1-z)^2\bigr ]\,
K_1(\varepsilon r)\,\varepsilon\,
\int \dfr p_T\,p_T^2\,J_1(p_T r)\Psi_{V_T}(p_T,z)\,.
\eeqn
%------------------------------------------------------
%

The spatial part of the LF quarkonium wave function (\ref{vm-wf}), $\Psi_{V}(r,z)$, can be found
in the framework of potential approach. As the first step, we start with determination of a quarkonium wave function $\Psi_{nlm}(\vec\rho)$ in the $Q\bar Q$ rest frame. In the considering 
case of $S$-wave production, one can safely represent $\Psi_{nlm}(\vec\rho)$ in the following factorized form,
%
%------------------------------------------------------
\beqn
\Psi_{nlm}(\vec\rho)=\Psi_{nl}(\rho)\cdot Y_{lm}(\theta,\varphi)\,,
\label{nrel-wf}
\eeqn
%------------------------------------------------------
%
where functions $\Psi_{nl}(\rho)$ and $Y_{lm}(\theta,\varphi)$ are the radial and orbital parts of the wave function, respectively. The former part depends on the three-dimensional $Q\bar Q$ separation $\vec\rho$ and has been obtained by solving the Schroedinger equation for different realistic potentials $V_{Q\bar Q}(\rho)$ for quark and antiquark interaction according to the formalism of Refs.~\cite{yura-00,sr-18,sr-19}. So far, there is no unambiguous way to relate the radial wave function of the lowest Fock component $|Q\bar Q\ra$ in the $Q\bar Q$ rest frame, $\Psi_{nl}(\rho)$, with its spatial wave function in the LF frame, $\Psi_{V}(r,z)$, by a simple Lorentz boost transformation. Usually, one employs a particularly simple recipe found in Ref.~\cite{terentev} and described in detail in Refs.~\cite{jan-96,yura-00,sr-18,sr-19}. As a result, one ends up with the momentum-space spatial quarkonium wave function in the LF frame $\Psi_V(p_T,z)$ which readily enters Eqs.~(\ref{PsiQQ}), (\ref{AL}) and (\ref{AT}). By Fourier transformation (\ref{PsiV_r-to-pT}), one obtains the corresponding impact-parameter spatial wave function $\Psi_V(r,z)$ to be used in Eq.~(\ref{vm-wf}).

The effect of Melosh spin rotation in the total cross sections of $\gamma^*\,p\rightarrow \Jpsi (\psip, \Y, \Yp, \Ypp)\,p$ processes is not connected to a particular form of the interaction potential $V_{Q\bar Q}(\rho)$. We also assume for simplicity that the $D$-wave effect comes only through the structure of the $V\to Q\bar Q$ transition, and is not related to the potential as well. Consequently, following the results of Refs.~\cite{sr-18,sr-19}, in the present study we adopt the same Buchm\"uller-Tye (BT) \cite{bt-80} potential and compare our calculations also with the predictions based on the power-like (POW) potential \cite{barik-80} in the case of charmonium production.

As was mentioned above, the essential ingredient of the production amplitude (\ref{prod-amp}) in the dipole picture is the universal phenomenological dipole cross section $\sigma_{Q\bar Q}(x,r)$ that allows to describe in an uniform way various high-energy processes, both inclusive and diffractive. It represents the interaction of a $Q\bar Q$ dipole of a transverse separation $\vec r$ with the proton target (at a given Bjorken $x$) and, consequently, the magnitudes of $\sigma_{Q\bar Q}$ at different values $\vec r$ are the eigenvalues of the elastic amplitude operator. Due to flavor universality of the QCD coupling, the dipole cross section is also typically considered to be flavor-independent. Another pronounced feature is known as the color transparency: $\sigma_{Q\bar Q}(r) \propto r^2$ for $r\!\!\to\!0$.

So far, the dipole cross section cannot be predicted reliably from the first principles due to poorly known higher-order pQCD corrections and nonperturbative effects. Therefore, we are obliged to use only its phenomenological parametrizations obtained from the fit of the high-precision DIS data available from HERA collider. Although many different parametrizations for $\sigma_{Q\bar Q}(x,r)$ can be found in the literature, for our purposes here we use only two of them following the results of Ref.~\cite{sr-18}. The first, simplest but phenomenologically very successful parametrization conventionally denoted as GBW has been suggested in Ref.~\cite{gbw}. The second parametrization denoted as KST has been proposed in Ref.~\cite{kst-99}. The choice of these models is also motivated by our early observation that they do not require any skewness corrections as those are assumed to already be effectively incorporated via the corresponding fits to the HERA data. A more detailed description of both models and the skewness corrections can be found in Ref.~\cite{sr-18} and will not be repeated further on in this work.

As was already emphasized in Ref.~\cite{sr-18}, one can treat the variations in predictions based upon the GBW \cite{gbw} and KST \cite{kst-99} models for the dipole cross section as a good measure of the underlined theoretical uncertainties. Another choice among the available dipole parameterizations has practically no impact on conclusions of this work that concern the relative onset of $D$-wave and spin effects in exclusive photo- and electroproduction of heavy quarkonia.

%
%
%
%===================================
\section{Numerical results: the impact of $D$-wave and spin effects}
\label{Sec:results}
%===================================
%
%
%
%
%%%%%%%%%%%%%%%%%%%%%%%%%%%%%%%%%%%%%%%%%%%%%%%%%%%%%%%%%%%%%%%%%%%%%%%%%%%%%%%%
\begin{figure}[!t]
  \includegraphics[scale=1.2]{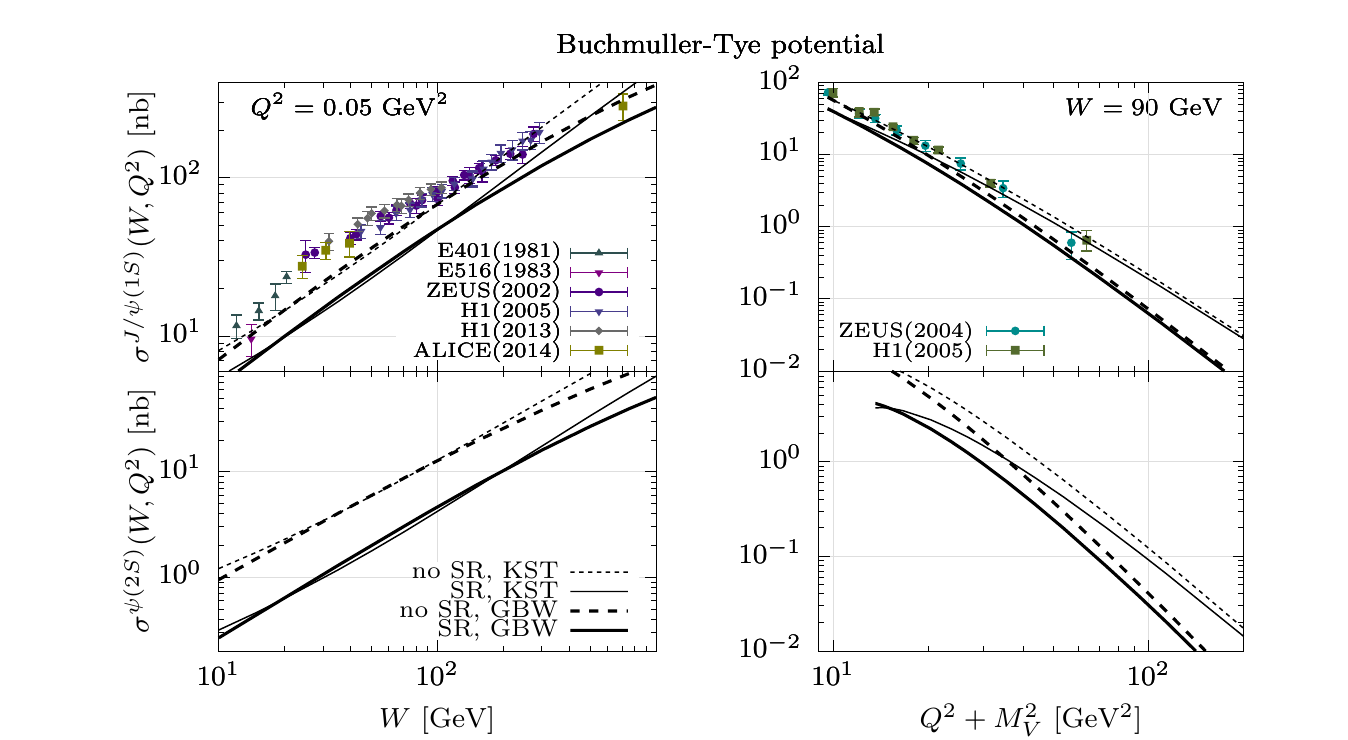}\centering  
\caption{\label{fig:bt-psi}  %(color online)
 %-------------------------------------
 Illustration of the impact of the Melosh spin rotation on the charmonium electroproduction cross sections obtained with the photon-like $V\to Q\bar Q$ vertex operator. {\bf Left panels:} Cross sections for $\gamma\,p\to \Jpsi(1S)\,p$ (left upper panel) and $\gamma\,p\to \psip(2S)\,p$ (left lower panel) processes with $Q^2 = 0.05\,\GeV^2$ as functions of $\gamma p$ collision c.m. energy $W$ (the experimental data are from Refs.~\cite{e401,e516,zeus-02,h1-05,h1-13,alice-14}). {\bf Right panels:} The same cross sections but as functions of the scale $Q^2 + M_{\Jpsi}^2$ for the $\Jpsi(1S)$ electroproduction (right upper box) and $Q^2 + M_{\psip}^2$ for the $\psip(2S)$ electroproduction (right lower box) at fixed $W=90\,\GeV$ (the data are from Refs.~\cite{zeus-04,h1-05}). Our results have been obtained using the wave functions of $\Jpsi(1S)$ and $\psip(2S)$ mesons generated by the BT potential \cite{bt-80}. The results without (scenario I) and with (scenario II) the Melosh spin transform are indicated by dashed and solid lines, respectively. Thin and thick lines represent the predictions obtained with KST \cite{kst-99} and GBW \cite{gbw} dipole parameterisations, respectively.
 %--------------------------------------
  }
\end{figure} 
%%%%%%%%%%%%%%%%%%%%%%%%%%%%%%%%%%%%%%%%%%%%%%%%%%%%%%%%%%%%%%%%%%%%%%%%%%%%%%%%
%

As the first step let us follow the common trend in the literature devoted to studies of exclusive quarkonium electroproduction (see e.g. Refs.~\cite{ryskin-92,brodsky-94,frankfurt-95,jan-94,jan-96,jan-98,jan-01}) and assume the same Lorentz structure for $V\rightarrow Q\bar Q$ and $\gamma^*\rightarrow Q\bar Q$ wave functions (scenario I, see above). For completeness, we also consider scenario II when the same structure of both transitions is imposed in the $Q\bar Q$ rest frame, in order to fix the relative weights of the $S$- and $D$-wave components. In one way or another, the quantitative role of the $D$-wave component remains unclear so far, and we wish to understand it better by comparing the results of the both scenarios I and II with each other and with the results of the previously studied ``$S$-wave only'' case (scenario III).

As was thoroughly described in the previous section, the calculations of the exclusive quarkonium electroproduction cross sections are performed in the framework of color dipole approach. Besides, in scenarios II and III starting from $S+D$-wave and $S$-wave structures, respectively, we incorporate also the effects of Melosh spin rotation representing the transformation of the spin-dependent components of the quarkonium wave function from the meson rest frame to the LF frame. Such analysis of the spin effects is performed here for the first time treating the photon-like structure of the $V\rightarrow Q\bar Q$ vertex and is complementary to our previous studies \cite{sr-18,sr-19} employing a different ($S$-wave only) and, hence, simpler structure of the vector-meson LF wave function. The corresponding numerical results have been obtained using Eq.~(\ref{final}) with amplitudes (\ref{AT}), (\ref{AL}).
%
%%%%%%%%%%%%%%%%%%%%%%%%%%%%%%%%%%%%%%%%%%%%%%%%%%%%%%%%%%%%%%%%%%%%%%%%%%%%%%%%
\begin{figure}[!h]
  \includegraphics[scale=1.2]{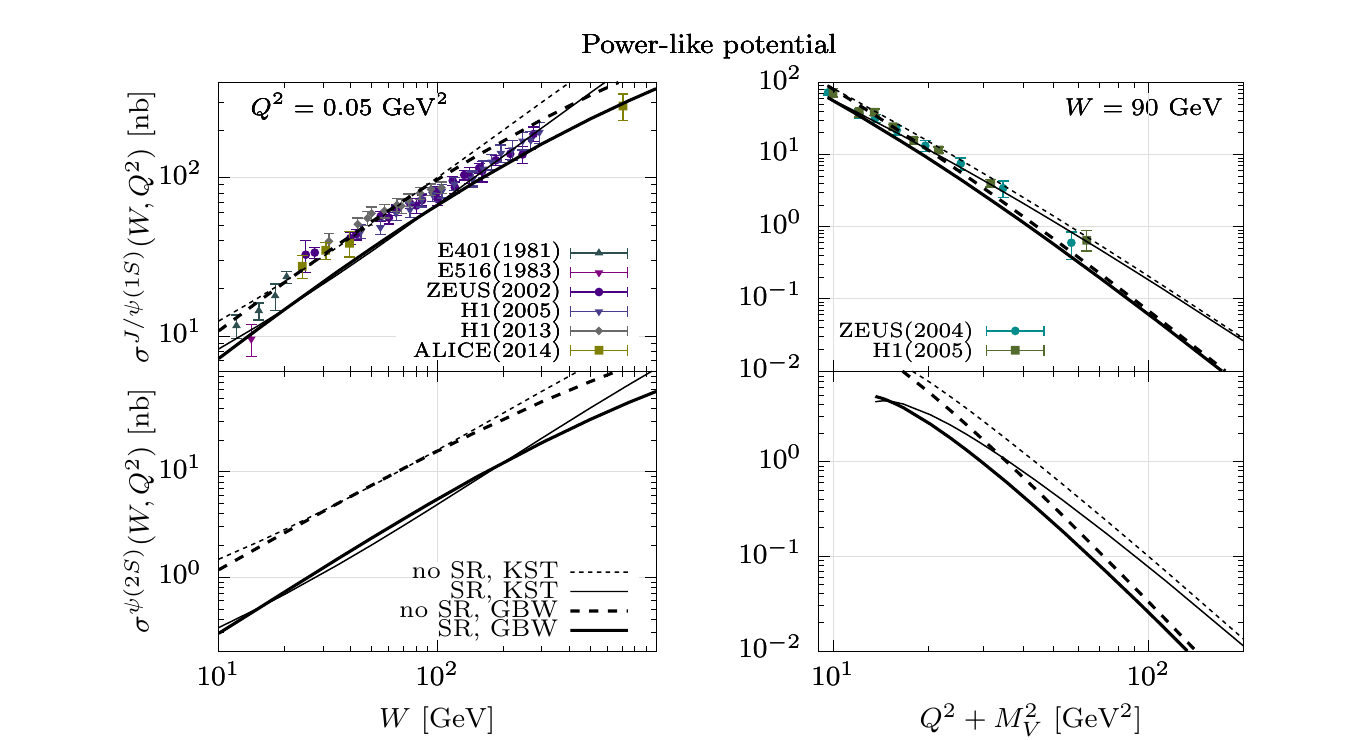}\centering  
\caption{\label{fig:pow-psi}  %(color online)
  %-------------------------------------
  The same as Fig.~\ref{fig:bt-psi}, except using the $\Jpsi(1S)$ and $\psip(2S)$ radial wave functions computed with 
  the power-like $Q\bar Q$ potential \cite{barik-80}.
  %-------------------------------------
  }
\end{figure}  
%%%%%%%%%%%%%%%%%%%%%%%%%%%%%%%%%%%%%%%%%%%%%%%%%%%%%%%%%%%%%%%%%%%%%%%%%%%%%%%%%
%
%
%%%%%%%%%%%%%%%%%%%%%%%%%%%%%%%%%%%%%%%%%%%%%%%%%%%%%%%%%%%%%%%%%%%%%%%%%%%%%%%%
\begin{figure}[tb]
		\begin{center}
		\includegraphics[width=.95\textwidth]{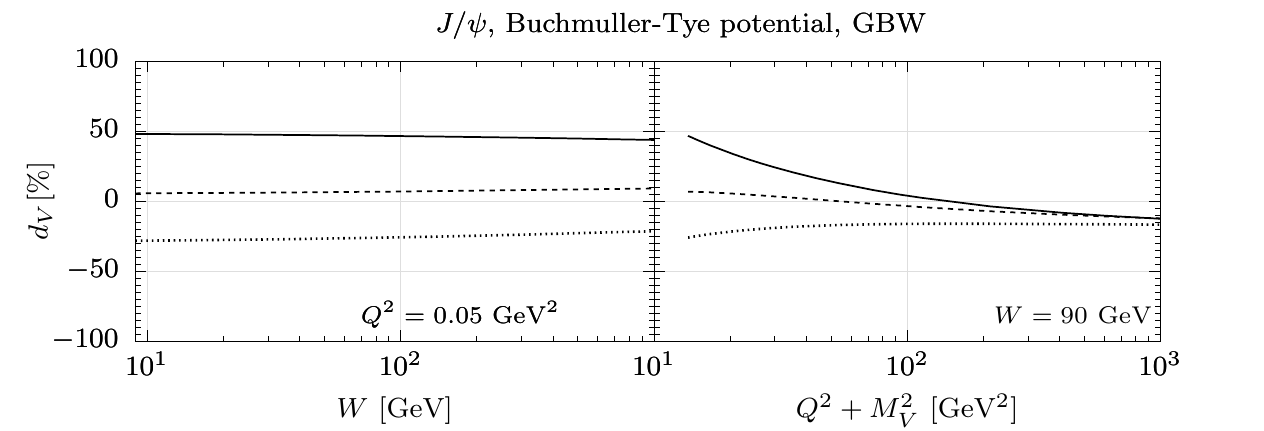}
		\includegraphics[width=.95\textwidth]{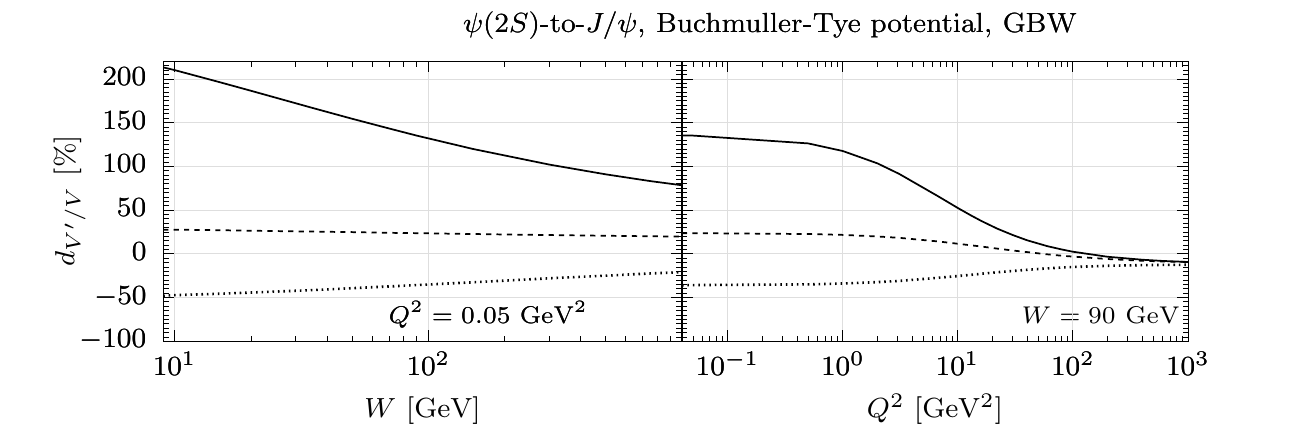}
		\end{center}
\caption{\label{fig:D-wave}
  %-------------------------------------
  Relative impact of the $D$-wave and Melosh spin transform on charmonium production 
  from a comparison of scenarios I, II, III and III$_{\rm no SR}$ (i.e. ``$S$-wave only'' case but without Melosh transform): electroproduction cross section for $\Jpsi(1S)$ (upper panels) and $\psip(2S)$-to-$\Jpsi(1S)$ ratio of electroproduction cross sections (lower panels). Solid lines represent the ratio (I - III$_{\rm no SR}$) / III$_{\rm no SR}$, dashed lines: (I - III) / III, and dotted lines: (II - III) / III. The results have been obtained using GBW \cite{gbw} dipole parameterisation and with the wave functions of $\Jpsi(1S)$ and $\psip(2S)$ mesons generated by the BT potential \cite{bt-80}.
  %-------------------------------------
  }
\end{figure}  
%%%%%%%%%%%%%%%%%%%%%%%%%%%%%%%%%%%%%%%%%%%%%%%%%%%%%%%%%%%%%%%%%%%%%%%%%%%%%%%%%
%

In Figs.~\ref{fig:bt-psi} and \ref{fig:pow-psi} we present the results for the total cross sections of $\gamma^*\,p\to \Jpsi\,p$ (upper panels) and $\gamma^*\,p\to \psip\,p$ (lower panels) processes and for the BT and POW potentials, respectively. The cross sections for $\Jpsi(1S)$ production are shown together with the available data as functions of $\gamma p$ collision energy $W$ and the hard scale of the process, $Q^2 + M_{\Jpsi}^2$. These calculations were performed by using the KST and GBW dipole parametrizations as shown by thin and thick lines, respectively.

Both Figs.~\ref{fig:bt-psi} and \ref{fig:pow-psi} clearly demonstrate a good agreement of the results with the available data when considering the scenario I.
%which is well-known result in the literature. 
In Fig.~\ref{fig:D-wave} (dashed lines, upper panels) we notice that scenarios I and III provide quantitatively very similar results for $\Jpsi(1S)$ production. However, for production of radially excited charmonium states, such as $\psip(2S)$, the $S+D$-wave result in the scenario I grows up to 25\% above the ``$S$-wave only'' result as is depicted by dashed lines in lower panels of Fig.~\ref{fig:D-wave}. 
Such a result demonstrates an enhanced role of the $D$-wave component in production of radially excited quarkonia as a consequence of the nodal structure of their radial wave functions.

\begin{figure}[!h]
  \includegraphics[scale=1.2]{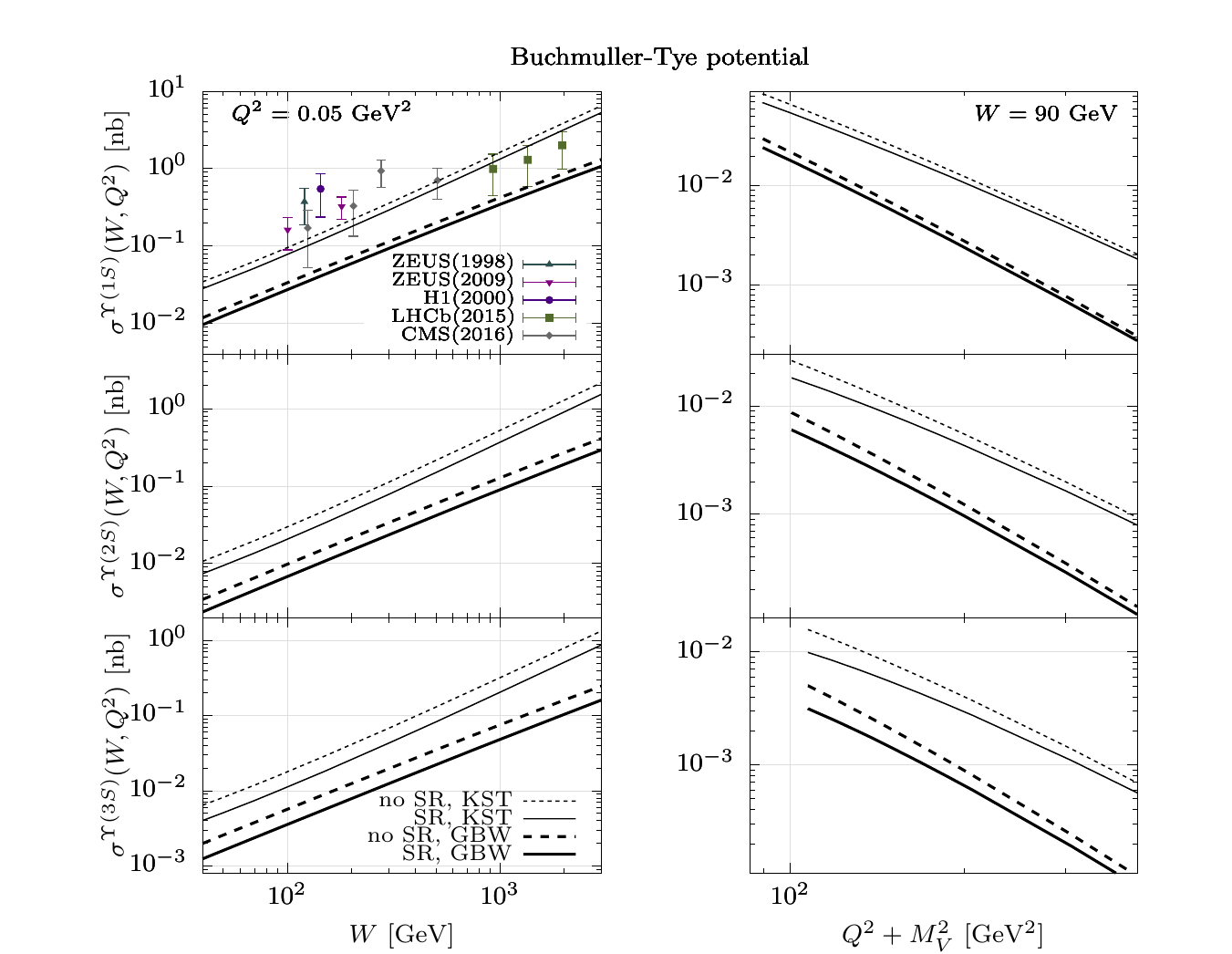}\centering  
\caption{\label{fig:bt-upsilon}
  %------------------------------------------------
  The same as Fig.~\ref{fig:bt-psi} but for the electroproduction of $\Y(1S)$ (upper panels), $\Yp(2S)$ (middle panels) and $\Ypp(3S)$ (lower panels) states. The experimental data for $\gamma\,p\to \Y\,p$ process are from Ref.~\cite{zeus-98,zeus-09,h1-00,lhcb-15,cms-16}. Here, the radial wave functions of $\Y(1S)$, $\Yp(2S)$ and $\Ypp(3S)$ states are generated by the BT potential \cite{bt-80}.
  %-------------------------------------------------
  }
  \end{figure}

The both Figs.~\ref{fig:bt-psi} and \ref{fig:pow-psi} also show that
%On the other hand, 
the scenario II (with the explicit Melosh spin transformation acted upon the photon-like $V\to Q\bar Q$ vertex in the meson rest frame) leads to a significant $20-30\,\%$ reduction of the photoproduction cross section for the $1S$ charmonium spoiling so its correspondence with the data (see also dotted lines in Fig.~\ref{fig:D-wave}, upper panels). This effect is even more dramatic in the case of $\psip(2S)$ production. A very similar trend can be seen when comparing the results for bottomonium production in its ground ($1S$) and excited ($2S$ and $3S$) states shown in Fig.~\ref{fig:bt-upsilon}. 
%
%%%%%%%%%%%%%%%%%%%%%%%%%%%%%%%%%%%%%%%%%%%%%%%%%%%%%%%%%%%%%%%%%%%%%%%%%%%%%%%%%%%%%%%
%%%%%%%%%%%%%%%%%%%%%%%%%%%%%%%%%%%%%%%%%%%%%%%%%%%%%%%%%%%%%%%%%%%%%%%%%%%%%%%%%%%%%%%%
%
%
%%%%%%%%%%%%%%%%%%%%%%%%%%%%%%%%%%%%%%%%%%%%%%%%%%%%%%%%%%%%%%%%%%%%%%%%%%%%%%%%
\begin{figure}[!t]
  \includegraphics[scale=1.16]{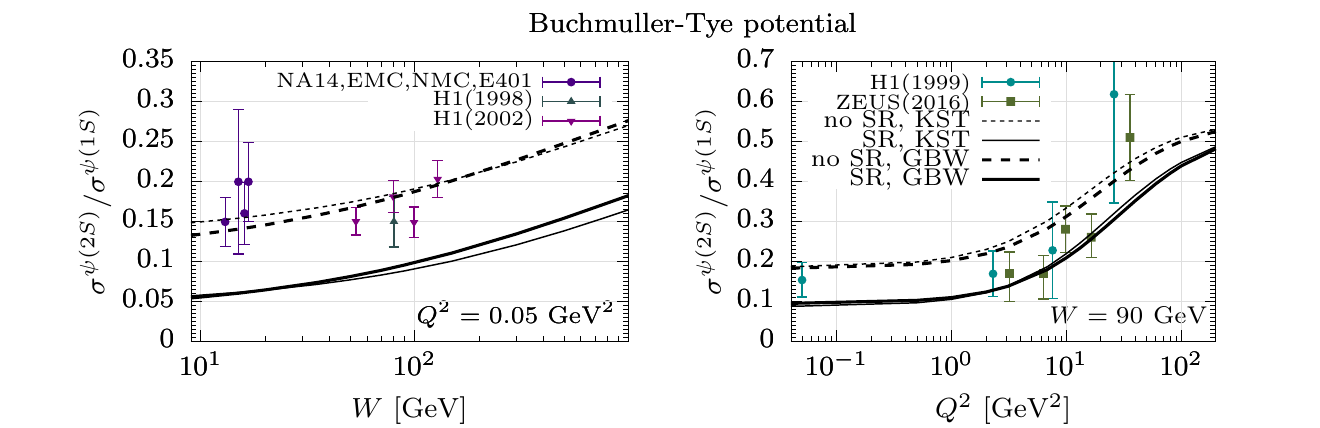}\centering  
  \vspace*{0.10cm}
\caption{\label{fig:bt-rat-psi}
  %----------------------------------------------------------
  {\bf Left panel:} The $\psip(2S)$-to-$\Jpsi(1S)$ cross section ratio as a function $\gamma p$ collision c.m. energy $W$ at fixed $Q^2 = 0.05\,\GeV^2$. 
  {\bf Right panel:} The same quantity but as a function of $Q^2$ at fixed $W = 90\,\GeV$.
  The predictions are compared to the experimental data taken from Refs.~\cite{na14-rat,emc-rat,nmc-rat,e401-rat,h1-98,h1-02,h1-99,zeus-16}. For more details, see the caption of Fig.~\ref{fig:bt-psi}.
  %----------------------------------------------------------
  }
\end{figure}  
%%%%%%%%%%%%%%%%%%%%%%%%%%%%%%%%%%%%%%%%%%%%%%%%%%%%%%%%%%%%%%%%%%%%%%%%%%%%%%%%%
%
%
%%%%%%%%%%%%%%%%%%%%%%%%%%%%%%%%%%%%%%%%%%%%%%%%%%%%%%%%%%%%%%%%%%%%%%%%%%%%%%%%%%%%%%%%%%%%%%%
\begin{figure}[!t]
  \vspace*{3.0cm}
  \includegraphics[scale=1.2]{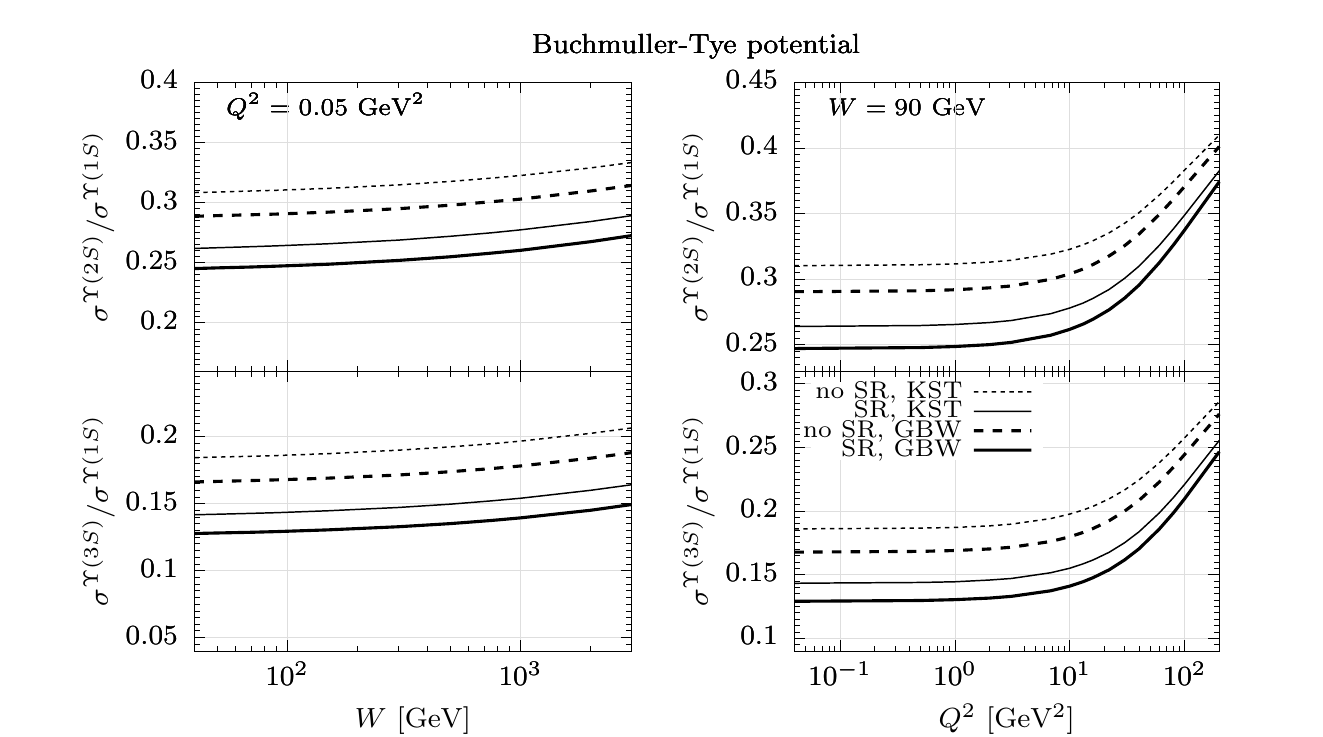}\centering  
  \caption{\label{fig:bt-rat-ups}  %(color online)
%-------------------------------------
The same as in Fig.~\ref{fig:bt-rat-psi} but for $\Yp(2S)$-to-$\Y(1S)$ (upper panels) and $\Ypp(3S)$-to-$\Y(1S)$ (lower panels) cross section ratios obtained using the BT potential \cite{bt-80}.
%-------------------------------------
} 
\end{figure}  
%%%%%%%%%%%%%%%%%%%%%%%%%%%%%%%%%%%%%%%%%%%%%%%%%%%%%%%%%%%%%%%%%%%%%%%%%%%%%%%%%%%%%%%%%%%%%%%%
%

Indeed, we would like to stress that the magnitude of spin effects is correlated with a structure of the $V\rightarrow Q\bar Q$ transition. 
As was pointed out in Refs.~\cite{sr-18,sr-19}, starting from the pure $S$-wave $V\to Q\bar Q$ transition in the meson rest frame, the explicit Melosh spin rotation (scenario III) in turn increases significantly the corresponding cross sections and as a result provides a reasonably good description of the experimental data. The magnitude of spin effects can be thus noticed directly by comparing the solid and dashed lines in Fig.~\ref{fig:D-wave}. 

However, the presence of the $D$-wave contribution modifies the onset of spin effects as one can notice by comparing dashed and dotted lines in Fig.~\ref{fig:D-wave}.
The $D$-wave component, when combined with the effect of the Melosh spin transform (scenario II), leads to a very strong (negative) effect in the quarkonium production observables. While the Melosh transform applied to the ``$S$-wave only'' enhances the cross section, the presence of the $D$-term effectively reverses this trend and instead dumps the cross section. As an interesting observation here, the numerical effect of the $D$-wave contribution is roughly close to the effect of the Melosh transform in the ``$S$-wave only'' case, but working in opposite directions (cf. solid and dotted lines in Fig.~\ref{fig:D-wave} with respect to dashed lines).

In comparison to production of 1S-state heavy charmonia,
the nodal structure of wave functions for the radially-excited states causes a much stronger onset of the spin effects in electroproduction of $\psip(2S)$
as is depicted in lower panels of Figs.~\ref{fig:bt-psi} and \ref{fig:pow-psi}. The lower panels of Fig.~\ref{fig:D-wave} demonstrate the relative magnitude of $D$-wave effect (with and without the spin rotation) in the $\psip(2S)$-to-$\Jpsi(1S)$ ratio of the electroproduction cross sections and its behaviour with energy and $Q^2$ for different scenarios. The corresponding ratio computed for scenarios I and II is compared to the HERA data in Fig.~\ref{fig:bt-rat-psi} in the left and right panel where it is shown as a function of $W$ and $Q^2$, respectively.

As was already mentioned in Sect.~\ref{Sec:Intro} and also discussed in Ref.~\cite{sr-18} the scanning radius given by Eq.~(\ref{scan-rad}) controls the universality between different cross sections. Consequently, one should expect the same magnitude of electroproduction cross sections as well as the same onset of $D$-wave and spin effects for $\gamma^*\,p\to \Jpsi(1S)\,p$ process at some scale $Q^2(\Jpsi)$, and for $\Y(1S)$ photoproduction. Such value $Q^2(\Jpsi)\sim 130\div 140\,\GeV^2$ has been estimated in Ref.~\cite{sr-18} from Eq.~(\ref{scan-rad}). 

Figs.~\ref{fig:bt-psi}, \ref{fig:pow-psi} and \ref{fig:D-wave} (right panels) demonstrate a gradual disappearance of the $D$-wave and spin rotation effects with $Q^2$ and their very weak onset at large scale $Q^2(\Jpsi)$.  Thus, following the scanning phenomenon, one expects such a weak onset of $D$-wave and spin effects also in $\Y(1S)$ photoproduction as is shown in Fig.~\ref{fig:bt-upsilon}. Here, in contrast to real photoproduction of $\Jpsi(1S)$, such marginal spin effects does not spoil an agreement with available data as is shown in upper left panel of Fig.~\ref{fig:bt-upsilon}.

The Fig.~\ref{fig:bt-upsilon} shows also our results for the energy and scale dependencies of the elastic (real and virtual) photoproduction cross sections for radially excited bottomonium states $\Yp(2S)$ and $\Ypp(3S)$. A rough measure of theoretical uncertainties is indicated by typical differences in numerical results corresponding to the GBW and KST dipole parametrizations. The existence of nodes in $\Yp(2S)$ and $\Ypp(3S)$ radial wave functions leads to a stronger onset of both $D$-wave and spin rotation effects compared to production of the $\Y(1S)$ state as one can see in Fig.~\ref{fig:bt-rat-ups}.

%
% 
%  
%=========================
\section{Conclusions}
\label{Sec:conclusions}
%=========================
%  
%
%

The present analysis represents a natural continuation of our previous studies \cite{sr-18,sr-19} of the Melosh spin transformation in elastic electroproduction of $S$-wave heavy quarkonia, $\gamma^*\,p\to V\,p$, where the following states $V = \Jpsi(1S), \psip(2S), \Y(1S), \Yp(2S)$ and $\Ypp(3S)$ have been considered. Here, we are primarily focused on phenomenological implications of the combined effect of Melosh spin rotation and the $D$-wave contribution which is associated with the widely-used photon-like structure of the quarkonium wave function. While the latter is directly postulated in the light-front (or infinite-momentum) frame, when transformed to the quarkonium rest frame, it contains an unknown $D$-wave term which has an unclear quantitative role in calculations of corresponding observables, especially, for radially-excited states whose yields are highly sensitive to the nodal structures in their radial wave functions. Indeed, the $D$-wave contribution in such a standard but unjustified approach cannot be consistently obtained starting from a realistic interquark potential, while an analysis of its phenomenological implications is still missing in the literature.

In order to compute the electroproduction cross section for heavy quarkonia using the dipole model representation for the factorized light-front transversely (\ref{AT}) and longitudinally (\ref{AL}) polarized production amplitudes, one needs to correctly identify several important ingredients. One such ingredient is the light-front photon wave function for the leading-order Fock $\gamma^\star \to Q\bar Q$ fluctuation, another is the light-front vector meson wave function $\Psi_V(r,z)$ for the $S$-wave heavy quarkonium, and, finally, the universal dipole cross section $\sigma_{Q\bar Q}(x,r)$.

In Refs.~\cite{sr-18,sr-19}, we have used a specific structure of the quarkonium wave function corresponding to ``$S$-wave-only'' contribution in the $Q\bar Q$ rest frame. In the present work, we have derived for the first time the formula (\ref{final}) with amplitudes (\ref{AT}) and (\ref{AL}) for electroproduction of $T$ and $L$ polarized quarkonia. Here we have included the spin rotation effects in the case of a massive photon-like structure of the $V\to Q\bar Q$ light-front wave function, readily imposed in the $Q\bar Q$ rest frame. The latter is motivated by the fact that the first non-derivative term in the spin-dependent
part of such a photon-like wave function takes exactly the same form as the ``$S$-wave-only'' 
spin-dependent component already analysed in our previous papers ~\cite{sr-18,sr-19}. The other terms containing derivatives in such an ansatz are then attributed to the $D$-wave effect, thus, enabling us to study its quantitative role. 

The radial wave functions are straightforwardly computed in the potential approach by solving the Schroedinger equation for the realistic Buchm\"uller-Tye and power-like interquark interaction potentials.
Then, both spin-dependent and radial wave functions are boosted to the light-front frame by using the Melosh spin and Terent'ev transforms. Our current study is thus complementary to our previous analysis \cite{sr-18,sr-19} indirectly showing an interplay of the Lorentz structure of the quarkonium transition vertex and the Melosh spin rotation. Our results are compared to the HERA data on photoproduction of the ground as well as the radially-excited heavy quarkonium states. Besides, our results are compared to those obtained with the ``$S$-wave-only'' structure of the quarkonium wave function in order to analyse the combined effect of the $D$-wave and Melosh transform as well as with the standard photon-like structure of the quarkonium wave function postulated in the light-front frame.

Our observations are the following. \\
(i)
The onset of the interplaying $D$-wave and spin rotation effects is strongly correlated with a structure of the quarkonium vertex. As was shown in our previous analyses \cite{sr-18,sr-19} based on a distinct structure of the $V\to Q\bar Q$ and $\gamma^*\rightarrow Q\bar Q$ wave functions, the Melosh transformation enhances substantially the magnitude of corresponding photo- and electroproduction cross sections. However, in the case of a photon-like $V\to Q\bar Q$ transition postulated in the $Q\bar Q$ rest frame the spin effects cause a notable reduction of quarkonium production yields thus spoiling a reasonable agreement with the experimental data. Such an opposite response of both models for the quarkonium wave function is due to a dramatic role of the $D$-wave component whose numerical impact on the observables appears to be significant.
\\
(ii) 
The scanning phenomenon, represented by Eq.~(\ref{scan-rad}), leads to universal properties in production of various quarkonium states. Consequently, we found that the $D$-wave and spin effects are very similar in the charmonium and bottomonium cross sections at a fixed value of the hard scale $Q^2+M_V^2$.
\\
(iii)
In the case of radially-excited states, postulating the photon-like Lorentz structure of the quarkonium wave function in both the light-front frame and $Q\bar Q$ rest frame, our analysis shows that the correlated onset of $D$-wave and spin effects is stronger than in production of ground state $1S$ quarkonia due to a nodal structure of corresponding radial wave functions. Again, such a result stimulates a question about the correct structure of the heavy quarkonium wave functions which has to be taken into consideration when describing the recent sets of quarkonia photoproduction data coming from the LHC.
\\
(iv)
In the case of $\Jpsi(1S)$ electroproduction, the $D$-wave and spin rotation effects get gradually reduced with $Q^2$, and according to the scanning phenomenon we find a rather weak onset of these effects as well as in photoproduction of the $\Y(1S)$ state.
\\
(v)
The nodal structure of the radial wave function for $\Yp(2S)$ and $\Ypp(3S)$ states leads to a stronger combined onset of the $D$-wave and spin rotation effects than that in eletroproduction of $1S$ bottomonia.

In summary, an underestimation of the importance of the $D$-wave effects, considering the photon-like structure of the $V\rightarrow Q\bar Q$ transition in photo- and electroproduction of heavy quarkonia, may lead to serious problems with correct interpretation of the experimental data. For example, such a structure of the quarkonium vertex has been conventionally imposed in the light-front frame as the one that leads to viable predictions when compared to the HERA data and thus is frequently used in the ongoing studies (see e.g. Ref.~\cite{sr-19} and references therein). However, this is done without any physical justification and quantification of the $D$-wave component implicitly present in such results. The corresponding role of the $D$-wave contribution (combined with spin effects), which is not justified by any nonrelativistic $Q$-$\bar Q$ interaction potential, has been indirectly estimated by comparing predictions for production observables with those based on the ``$S$-wave-only'' structure of quarkonium wave functions in the $Q\bar Q$ rest frame.

Finally, it is worth mentioning that more precise data on quarkonium production in ultra-peripheral and heavy-ion collisions at RHIC and the LHC are certainly needed for a better understanding of the quarkonium formation dynamics. Moreover, the planed measurements at the future electron-ion colliders may shed more light on the onset of spin effects which can be considered as an important new probe for the underlying structure of the quarkonium wave function. Here the electroproduction of radially excited heavy quarkonia is even more effective for solving the problem associated with the structure of $V\rightarrow Q\bar Q$ transition. In comparison to electroproduction of $1S$ heavy quarkonium states, the nodal structure of their radial wave functions leads to a significantly higher sensitivity of our predictions for the corresponding cross sections to the quarkonium vertex structure and consequently to the onset of the Melosh spin effects.

%
%
%
%===========================
\section*{Acknowledgements}
%===========================
%
%
%

J.N. work was partially supported by grants LTC17038 and LTT18002 of the Ministry of Education, Youth and Sports of the Czech Republic, by the project of the European Regional Development Fund CZ02.1.01/0.0/0.0/16\_019/0000778, and by the Slovak Funding Agency, Grant 2/0007/18. 
R.P.~is supported in part by the Swedish Research Council grants, contract numbers 621-2013-4287 and 2016-05996, by the Ministry of Education, Youth and  Sports of the Czech Republic, project LT17018, as well as by the European Research Council (ERC) under the European Union's Horizon 2020 research and innovation programme (grant agreement No 668679). The work of M.K. was supported in part by the Conicyt Fondecyt grant Postdoctorado N.3180085 (Chile) and by the grant 18-07846Y of the Czech Science Foundation (GACR). The work has been performed in the framework of COST Action CA15213 ``Theory of hot matter and relativistic heavy-ion collisions'' (THOR).

%
%=====================================================================================
%

%===========================
\bibliographystyle{unsrt}

\end{document}